\documentclass[10pt]{amsart}

\usepackage{amsmath,amsfonts,amssymb}
\usepackage{graphicx,verbatim}
\usepackage[numbers,square,sort&compress]{natbib}
\usepackage[colorinlistoftodos]{todonotes}
\usepackage{times}
\usepackage{xspace}
\usepackage{xypic}
\usepackage{color}
\usepackage{a4wide}
\usepackage{subfig}
\usepackage{enumerate}

\usepackage{tikz}
\usetikzlibrary{arrows}
\usetikzlibrary{decorations.markings}
\usetikzlibrary{decorations.pathreplacing}
\usetikzlibrary{patterns}
\usepackage{tkz-graph}
\usetikzlibrary{shapes.geometric}

\theoremstyle{plain}

  \newtheorem{defn-prop}[thm]{Definition-Proposition}

\theoremstyle{definition}

\newcommand{\longto}{\longrightarrow}

\def\F{\mathbb{F}}

\def\And{\wedge}
\def\Or{\vee}

\tikzset{
  activ/.style={
    decoration={markings,mark=at position 1 with {\arrow[scale=2]{stealth}
      }
    }, 
    postaction={decorate}
  }
}
\tikzset{
    inhib/.style={
      shorten >=#1,
      decoration={
        markings,
        mark={
          at position 1
          with {
            \draw[fill] circle [radius=#1];
          }
        }
      },
      postaction=decorate
    },
    inhib/.default=1.75pt
}

\begin{document}

\title{Synchrony in a Boolean network model of the {\sc l}-arabinose operon in Escherichia coli}

\author[A.~Jenkins]{Andy Jenkins} \address{Department of
  Mathematics \\ University of Georgia \\ Athens, GA 30602, USA}
\email{lee.jenkins25@uga.edu}

\author[M.~Macauley]{Matthew Macauley} \address{Department of
  Mathematical Sciences \\ Clemson University \\ Clemson, SC 29634-0975, USA}
\email{macaule@clemson.edu}

\thanks{Partially supported by a National Science Foundation grant
  (DMS-1211691) and a Simons Foundation collaboration grant for
  mathematicians (Award \#358242).}



\begin{abstract}
The lactose operon in \emph{Escherichia coli} was the first known gene regulatory network, and it is frequently used as a prototype for new modeling paradigms. Historically, many of these modeling frameworks use differential equations. More recently, Stigler and Veliz-Cuba proposed a Boolean network model that captures the bistability of the system and all of the biological steady states. In this paper, we model the well-known arabinose operon in E.~coli with a Boolean network. This has several complex features not found in the \emph{lac} operon, such as a protein that is both an activator and repressor, a DNA looping mechanism for gene repression, and the lack of inducer exclusion by glucose. For $11$ out of $12$ choices of initial conditions, we use~ computational algebra and Sage to verify that the state space contains a single fixed point that correctly matches the biology. The final initial condition, medium levels of arabinose and no glucose, successfully predicts the system's bistability. Finally, we compare the state space under synchronous and asynchronous update, and see that the former has several artificial cycles that go away under a general asynchronous update. 
\end{abstract}

\maketitle

\section{Introduction}

The regulation of gene expression is essential for the maintenance of homeostasis within an organism. As such, the ability to predict which genes are expressed and which are silenced based on the cellular environment is of utmost importance to molecular biologists. Mathematical models of gene regulatory networks have classically been framed as systems of differential equations that are derived from mass-action kinetics. The result is a complex system of equations that cannot be solved explicitly, and involves many experimentally determined rate constants. In many cases, estimates of these constants in the literature may vary by orders of magnitude. The end result is a model which is quantitative by nature, but can only really be analyzed qualitatively, even if it can be solved numerically. These continuous models are still useful for understanding the mechanisms of regulation, but like any model, they have their strengths and weaknesses. An alternative approach is to use discrete models, such as logical models or Boolean networks \cite{thomas1990biological}. These too have their drawbacks, but they have been shown to be useful tools when one is interested in analyzing the global-level dynamics of the system. The interested reader can consult \cite{dejong2002modeling} for a nice survey in modeling gene regulatory networks, including both continuous and discrete models. The lactose operon in \emph{Escherichia coli} was the first gene regulatory network discovered, and Francois Jacob and Jacques Monod won a Nobel Prize in 1965 for their work on it \cite{jacob1960l'operon}. It is an example of an inducible operon under negative control, and a Boolean network model was proposed by Veliz-Cuba and Stigler in \cite{veliz-cuba2011boolean}. In this paper, we propose a Boolean model for the {\sc l}-arabinose operon. While this operon is also used by \emph{E. coli} to regulate sugar metabolism, it contains several unique biological features such as a protein that can act both as a positive inducible control mechanism, and as a negative control mechanism by forming a DNA loop to block transcription. As far as we know, our model is the first Boolean network model of a system that uses DNA looping to regulate transcription. 

We use computational algebra and Sage to analyze our model, which has $9$ variables and $4$ parameters. This leads to $12$ possible initial conditions, and our  analysis shows that for all of these, our model accurately predicts the biological behavior of the operon and also provides insight into interactions within the network. Additionally, when there are medium levels of arabinose, the model predicts the observed bistability of the system. However, our model under synchronous update contains a few artificial cycles, but these go away when we look at a general asynchronous update scheme. This should serve as a cautionary tale as to how certain artifacts can arise from a synchronous update in a molecular network model.

This paper is organized as follows: in Section~\ref{sec:ara-operon}, we begin with a review of the biology of the \emph{ara} operon, and we follow that with a review of Boolean and logical models. Finally, we propose our model and justify the choice of functions. In Section~\ref{sec:analysis}, we convert our model into polynomials in $\F_2[x_1,\dots,x_n]$ and we show how to use Sage to compute Gro\"obner bases to find the fixed points. We also use software such as ADAM \cite{hinkelmann2011adam}, TURING \cite{turing}, and GINsim \cite{chaouiya2012logical} to verify that no other limit cycles exist. We explain how our model captures bistability, and discuss the dependence of the system dynamics on the whether the functions are applied synchronously or asynchronously. We conclude in Section~\ref{sec:conclusions} with a discussion of the results and future work.

\section{The {\sc l}-arabinose operon in E. coli}\label{sec:ara-operon}

\subsection{Biological background}

A core theme throughout molecular biology is the so-called ``central dogma'', which explains how information flows from genotype (the genetic code) to the phenotype (the observable physical characteristics of an organism) \cite{crick1970central}. Genetic information encoded in the nucleotides of the DNA strand undergoes transcription, in which RNA polymerase enzymes ``unzip'' the DNA and read it, producing a messenger RNA strand.  This mRNA then undergoes translation by ribosomes into an amino acid sequence known as a polypeptide, or protein. The functions of these proteins then lead to the phenotype. Clearly this process requires extensive regulation, as the desire for presence or absence of a protein, as well as its concentration, are dependent on the current state of the organism and its environmental conditions.  Regulation of gene expression can occur at any of the steps during the flow of information, but the regulation of transcription in prokaryotic organisms has been particularly well-studied, and most of the early understanding of genetic regulatory systems came from the study of operons.  An \emph{operon} is a collection of contiguous genes with related functions that are all transcribed together onto a single mRNA strand, along with two control sequences upstream of the genes.

The genes in an operon whose protein products perform some coordinated
function are known as structural genes. The two control sequences are
a \emph{promoter}, a region of DNA to which RNA polymerase binds to initiate
transcription, and an \emph{operator}, a sequence of nucleotides located
between the promoter and structural genes whose status determines
whether transcription will occur. The state of the operator and its
method of controlling gene expression can be categorized into one of
four types. Operons can be \emph{inducible}, which means that their
``default state'' is off, except when needed. The opposite is a
\emph{repressible} operon, which means that the genes are transcribed
unless there is a reason or need to turn them off. Additionally, the
method of regulation can be positive (an activator protein binds to
the operator to initiate transcription) or negative (a repressor
protein binds to the operator to prevent transcription).

One of the most well-studied prokaryotic organisms is \emph{E. coli}, a rod-shaped bacterium commonly found in the gut of mammals and birds. It has a short genome which has been fully sequenced, and its physiology is well-understood. Since \emph{E. coli} lives in the intestines, its nutrition depends on the diet of its host. For example, if the host consumes milk, then \emph{E. coli} is exposed to lactose, or milk sugar. Lactose can be used as an energy source, but gene products such as proteins and enzymes are needed to carry out tasks like transporting the molecules into the cell and then breaking them down. These gene products are not needed and therefore will not be produced if lactose is not available. The genes that regulate this make up the \emph{lac} operon, and it is an example of an inducible operon with negative control. It was the first genetic regulatory mechanism to be completely understood, and it is the prototypical example of a gene regulatory network. As such, it is standard material in introductory molecular biology classes, and is used as test cases for new modeling paradigms. 

Lactose is just one of many sugar molecules that \emph{E. coli} can use as an energy source. Another is the five-carbon sugar {\sc l}-arabinose (the {\sc l}- prefix stands for \emph{levorotation}, which describes its chirality), and its metabolism into carbon and energy is also controlled by the gene regulatory network called the \emph{ara} operon. Both lactose and arabinose are complex sugar molecules which need to be broken down before they can be used for energy. Glucose, a more simple molecule, is a preferred energy source because it does not have to be modified to enter the respiratory pathway. Thus, the \emph{lac} and \emph{ara} operons should be off if glucose is present. 

The \emph{ara} operon contains three structural genes, called \emph{araBAD} (that is, the adjacent \emph{ara} genes B, A, and D). Transcription of these genes is controlled by the promoter $p_{BAD}$. These genes code for thee enzymes which catalyze the following reactions needed for arabinose metabolism: the isomerase AraA converts {\sc l}-arabinose to {\sc l-}ribulose, the kinase AraB phosphorylates {\sc l}-ribulose to {\sc l}-ribulose-phosphate, and finally the epimerase AraD converts {\sc l}-ribulose-phosphate to {\sc d}-xylulose-phosphate, which then enters the pentose phosphate pathway. In addition to the enzymes needed to metabolize arabinose, proteins are needed to transport it through the cell membrane. This is controlled by two different transport systems, both located upstream from the ara operon. The \emph{araE} gene produces AraE, a membrane-bound protein that functions as a transporter in a low affinity transport system, and the \emph{araFGH} genes which produce three corresponding proteins that together form a high affinity transport system known as an ATP-binding cassette.

\begin{figure}
\hspace{-5mm}\includegraphics[width=.51\textwidth]{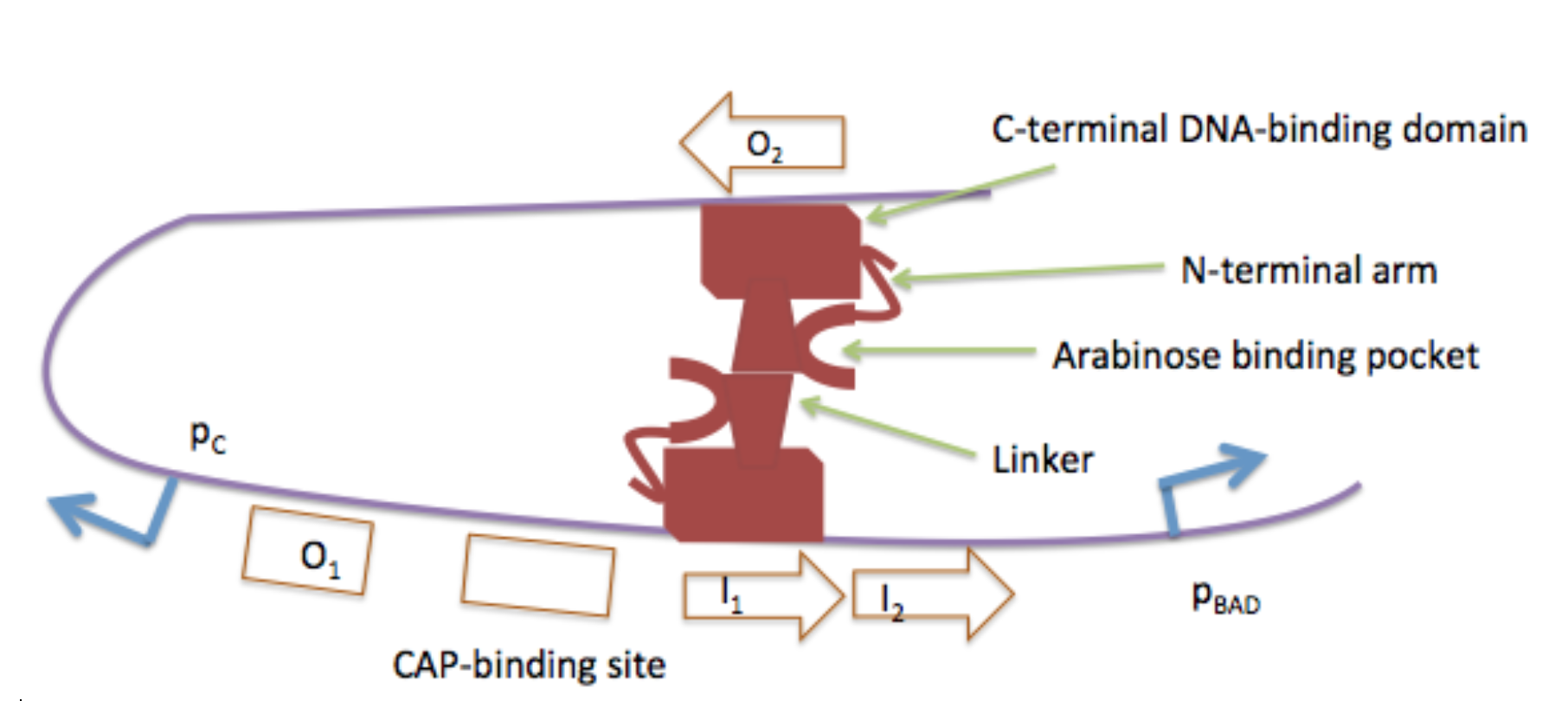}\includegraphics[width=.51\textwidth]{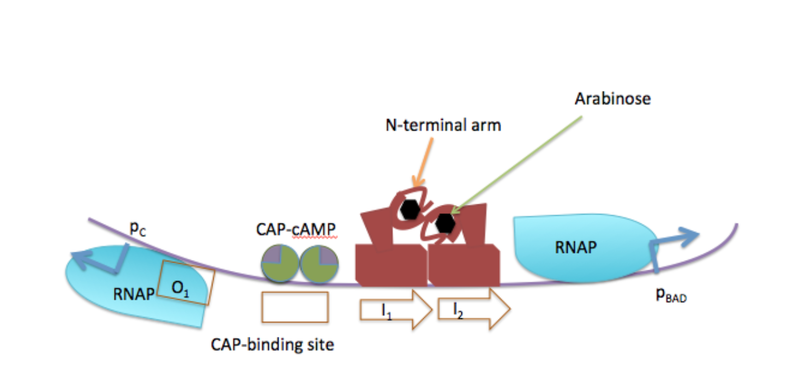}
\caption{Without arabinose (left), the AraC protein forms a dimer which bonds to the DNA strand in two places, causing the DNA to loop and block transcription. At right is the case of when arabinose is present. Images taken from Wikipedia.}\label{fig:ara-operon}
\end{figure}

Like the \emph{lac} operon, the \emph{ara} operon is an inducible operon, but it differs in that it has both positive and negative regulation. Specifically, the expression of the structural genes (\emph{araBAD}) is regulated by the protein AraC, which can function as either a repressor or an activator depending on the intracellular concentrations of arabinose. The cell maintains a small amount (approx. 20 molecules) of AraC at all times, which binds to two sites $I_1$, $O_2$ on the DNA strand that causes a DNA loop structure, as shown in the image on the left in Figure~\ref{fig:ara-operon}. This loop acts as a \emph{repressor} of transcription of both the \emph{araBAD} genes and the \emph{araC} gene by physically blocking RNA polymerase from attaching to either promoter sequence. This method of repression is quite different than the method used in the \emph{lac} operon, where a repressor protein binds directly onto the DNA strand to block transcription. However, the end result is the same. If extracellular arabinose is present, some molecules can be transported into the cell via passive transport. Once intracellular arabinose is available, these molecules will bind to the AraC protein and cause it to undergo a conformational change, which leads to it dissociating from the $O_2$ operator site and subsequently binding to the $I_2$ site. This is shown in the image on the right of Figure~\ref{fig:ara-operon}. In this arabinose-bound form, AraC now functions as an activator and induces binding of RNA polymerase to the $p_{BAD}$, $p_E$, $p_{FGH}$, and $p_C$ promoter regions. Transcription can now begin. In fact, the AraC protein was the first known positive regulator in an operon, which was verified in the early 1970s \cite{greenblatt1971arabinose}. Later experiments \cite{doyle1972induction}, \cite{ogden1980escherichia} uncovered the full structure of the operon; see \cite{schleif2000regulation} for a gentle survey article.

Since glucose is the preferred energy source to arabinose, it must play a
role in regulating the transcription of the \emph{araBAD} structural
genes, and it does this via a method called \emph{catabolite
  repression}. In addition to the arabinose-bound AraC protein, a
second activator known as the cyclic AMP catabolite activator protein
complex (or cAMP-CAP complex) must bind to the promoter region to
cause the DNA to undergo further conformational changes to allow the
binding of RNA polymerase. One can think of this complex as a ``key''
which is required for transcription to begin. The presence of glucose
inhibits the production of cAMP, which reduces the concentration
levels of the cAMP-CAP complex, thereby preventing transcription of
the \emph{ara} operon. The \emph{lac} operon not only uses catabolite repression, but also a form or repression called \emph{inducer exclusion}, where glucose binds to the transporter protein, causing a conformal change in preventing lactose from entering the cell. This mechanism is not used in the \emph{ara} operon \cite{saier1976sugar}, which actually makes modeling it more delicate.

\subsection{Boolean modeling background}

Molecular networks have been classically modeled using continuous methods, such as systems of differential equations. Examples of this for the \emph{lac} operon can be found in \cite{busenberg1985interaction}, \cite{yildirim2003feedback}, and \cite{yildirim2004dynamics}. The first Boolean network model of the \emph{lac operon} was published in 2011 \cite{veliz-cuba2011boolean}.

Boolean network models were first proposed in 1969 by theoretical biologist S.~Kauffman \cite{kauffman1969metabolic} as models of gene regulatory networks. In 1973, biologist Ren\'e Thomas published a discrete formalization of biological regulatory networks that are often called \emph{logical networks} \cite{thomas1973boolean}, and these continue to be studied and used in modeling today. The framework starts with a finite set $[n]:=\{1,\dots,n\}$ of nodes, each one having a state $x_i$ in a finite set $S_i=\{0,\dots,r_i-1\}$. The case of each $r_i=2$ is a \emph{Boolean network}. These are the most common, and henceforth, we will assume that $S_i=\F_2=\{0,1\}$. Each node also has an \emph{update function} $f_i\colon\F_2^n\to\F_2$ that updates its state based on the states of the other nodes. The update functions are then put together to generate the global system dynamics which can be encoded by a directed graph called the \emph{phase space}. There are several ways to do this, but in all of them, the vertex set is the set of global states $\F_2^n$, and the edges represent transitions between states. 

The two basic ways to generate the dynamics are by applying the update functions synchronously or asynchronously. In a synchronous update, the \emph{dynamical system map} $f\colon\F_2^n\to\F_2^n$ is defined by $f=(f_1,\dots,f_n)$. This defines a discrete-time dynamical system, and so the phase space consists of the $2^n$ edges of the form $(x,f(x))$, where $x\in\F_2^n$. Every node has one outgoing edge, and the periodic points are those on simple cycles; those of length $1$ are called \emph{fixed points}. One advantage of a synchronous update is the local functions and dynamical map are multivariate polynomials over $\F_2$, and so this opens the door to using the rich toolbox of computational algebra to study biological systems. This algebraic approach has been used extensively by R.~Laubenbacher and his collaborators; see \cite{laubenbacher2009computer}, \cite{veliz-cuba2010polynomial} for an overview. For example, when reverse-engineering a model from data, the model space consists of the sum of the vanishing ideal and a particular solution. Approaches to the model selection problem include computing an ideal and its primary composition \cite{laubenbacher2004computational}, or computing a Gr\"obner fan, selecting the cone with maximum weight, and then computing the Gr\"obner normal form of the corresponding Gr\"obner basis \cite{dimitrova2007grobner}. 

One major drawback to this approach is that molecular networks do not have a ``central clock'' that updates every component synchronously -- different gene products may have vastly different timescales, and there is always a degree of randomness involved. Therefore, many logical network models assume a \emph{general asynchronous update}. In this case, the phase space consists of the $n\cdot 2^n$ edges of the form $(x,f_i(x))$, though self-loops are frequently omitted. The time-evolution of the global system state can be described as a walk on the phase space. With probability $1$, a ``random walk'' will end up in a maximal strongly connected component that has no out-going edges. These could be fixed points or cycles (called \emph{cyclic attractors}), but they can also be more complicated subgraphs called \emph{complex attractors} \cite{thomas1990biological}. 

Regardless of whether the update functions are composed synchronously, asynchronously, or in some other method, the \emph{wiring diagram} is defined to be the graph with vertex set $[n]=\{1,\dots,n\}$ and a directed edge from $i$ to $j$ (or $x_i$ to $x_j$) if the function $f_j$ depends on $x_i$.  Such an edge is \emph{positive} if 
\[
f_j(x_1,\dots,x_{i-1},0,x_{i+1},\dots,x_n)\leq
f_j(x_1,\dots,x_{i-1},1,x_{i+1},\dots,x_n)\,
\]
and \emph{negative} if the inequality is reversed. Positive edges are typically denoted as \begin{tikzpicture}[scale=1,baseline=-.5ex]
  \node (A) at (0,0) {\footnotesize $x_i$};
  \node (B) at (1,0) {\footnotesize $x_j$};
  \draw[activ,shorten >= -2pt,shorten <= -2pt] (A) to (B);
\end{tikzpicture}, and negative edges as
\begin{tikzpicture}[scale=1,baseline=-.5ex]
  \node (A) at (0,0) {\footnotesize $x_i$};
  \node (B) at (1,0) {\footnotesize $x_j$};
  \draw[inhib,shorten >= -2pt,shorten <= -2pt] (A) to (B);
\end{tikzpicture} or
$\tikz[baseline=-0.5ex]{
      \node (xi){$x_i$}; \node (xj) [right of=xi] {$x_j$};
      \draw [-|] (xi) -- (xj);}$. Of course, edges can be neither positive nor negative, though most interactions in molecular networks are one or the other -- either activation or inhibition \cite{raeymaekers2002dynamics}.

\begin{figure}
  \tikzstyle{v} = [circle, fill=white,draw,inner sep=0pt, minimum size=4mm] 
  \tikzstyle{w} = [rectangle, fill=white,draw,inner sep=0pt, minimum size=6mm] 
  \tikzstyle{act} = [draw, -stealth]
  \begin{tikzpicture}
    \begin{scope}[shift={(4,0)},scale=1.5]
      \node (00) at (0,0) {\small $00$};
      \node (01) at (0,1) {\small $01$};
      \node (10) at (1,0) {\small $10$};
      \node (11) at (1,1) {\small $11$};
      \draw[activ] (00) to (01);
      \draw[activ] (00) to (10);
      \draw[activ] (11) to (01);
      \draw[activ] (11) to (10);
      \Loop[dist=.5cm,dir=WE](-.15,1);
      \Loop[dist=.5cm,dir=EA](1.15,0);
    \end{scope}
    \begin{scope}[shift={(0,0)},scale=1.5]
      \node (00) at (0,0) {\small $00$};
      \node (01) at (0,1) {\small $01$};
      \node (10) at (1,0) {\small $10$};
      \node (11) at (1,1) {\small $11$};
      \draw[activ] (00) to[bend left] (11);
      \draw[activ] (11) to[bend left] (00);
      \Loop[dist=.5cm,dir=WE](-.15,1);
      \Loop[dist=.5cm,dir=EA](1.15,0);
    \end{scope}
    \begin{scope}[shift={(-4.5,0)},scale=1.5]
      \draw[decorate,decoration={brace,amplitude=5pt}] (-2.6,.2) --  (-2.6,.8); 
      \node at (-1.75,.7) {$f_1(x_1,x_2)=\overline{x_2}$};
      \node at (-1.75,.3) {$f_2(x_1,x_2)=\overline{x_1}$};
      \node[v] (1) at (0,.5) {\small $1$};
      \node[v] (2) at (1.5,.5) {\small $2$};
      \draw[inhib] (1) to[bend left] (2);
      \draw[inhib] (2) to[bend left] (1);
    \end{scope}
  \end{tikzpicture}
\caption{A simple Boolean network and its 2-node wiring diagram. To the right is the phase space under a synchronous update, which contains two fixed points and a 2-cycle. On the far right is the phase space under general asynchronous update. This suggests that if the nodes are updated randomly, then the system will end up in one of the fixed points with probability $1$, and so the $2$-cycle is an artifact of synchrony.} \label{fig:2-node}
\end{figure}
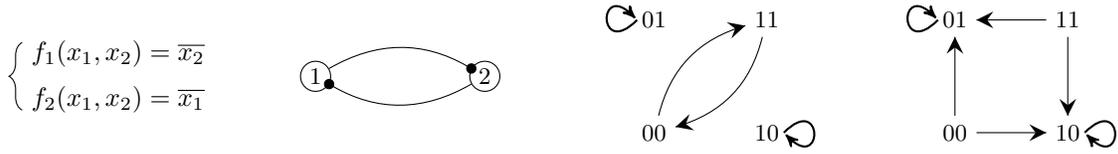

Early work on Boolean networks focused on models of well-known biological networks, as well as which network motifs in the wiring diagram can lead to common biological phenomenon. In 1981, Thomas conjectured \cite{thomas1981relation} that positive feedback loops are necessary for the dynamics to exhibit \emph{multi-stationarity} (having multiple fixed points) and negative feedback loops are necessary for cyclic attractors. In this context, a feedback loop is a directed cycle in the wiring diagram, and it is positive if it consists of an even number of negative edges, and odd otherwise. Biologically, a fixed point describes a biological steady state such as phenotype, and a cyclic attractor can describe a biological phenomenon like a cell cycle or homeostasis. Thomas' conjectures have been proven for both continuous systems such as differential equations (e.g., \cite{snoussi1998necessary}) as well as discrete system such as logical or Boolean networks (e.g., \cite{remy2008graphic}). It is easy to show that fixed points do not depend on the update order. Additionally, one can make a case that due to the robustness of biological systems, issues such as timing should not affect the long-term dynamics. However, a synchronous update can introduce artifacts into the phase space. A simple example of this is shown in Figure~\ref{fig:2-node}. Alternatively, non-determininstic versions of discrete models have been developed and studied. Examples include probabilistic Boolean networks \cite{murrugarra2012modeling} and stochastic discrete dynamical systems \cite{shmulevich2010probabilistic}.

Since the \emph{lac} operon was the first operon discovered and remains the most widely known, it is no surprise that it was the first operon to be modeled in a Boolean or logical network framework.\footnote{The \emph{trp} operon has been modeled in a discrete framework, but using Petri nets \cite{simao2005qualitative}.} This was done in 2011 by Veliz-Cuba and Stigler \cite{veliz-cuba2011boolean}. We will begin by summarizing their work here in a few sentences, and then carry out a similar undertaking for the \emph{ara} operon. In \cite{veliz-cuba2011boolean}, the authors proposed a model of the \emph{lac} operon with 10 Boolean variables that represent relevant gene products (mRNA, enzymes and proteins) and intracellular lactose. For most of these, one can think of $0$ as denoting ``absent'' or ``low (basal) concentration'', and $1$ as denoting ``present'' or ``high concentration''. A system state is thus a vector in $\F_2^{10}$, and they use a synchronous update $f=(f_1,\dots,f_{10})$ to get a dynamical system map  $f\colon \F_2^{10}\to\F_2^{10}$. Additionally, their model has three parameters: $G_e$, $L_e$, and $L_{em}$, which are treated as constants and appear in some of the update functions. Extracellular glucose is either present ($G_e=1$) or absent ($G_e=0$). Concentration levels of extracellular lactose can be one of three levels, which are described using two Boolean variables. The variable $L_e$ means ``high levels of extracellular lactose'' and $L_{em}$ means (at least) medium levels. Together, they describe high levels by $(L_e,L_{em})=(1,1)$, medium by $(L_e,L_{em})=(0,1)$, and low (basal) levels by $(L_e,L_{em})=(0,0)$. The fourth possibility, $(L_e,L_{em})=(1,0)$, is ignored because it is meaningless. This gives six possible values for the parameter vector $(L_e,L_{em},G_e)\in\F_2^3$, each representing an \emph{initial condition} of the model. 

For five of the six initial conditions, the authors verify that the phase space consists of exactly one basin of attraction (connected component) with one fixed point. Each of these fixed points either corresponds to the \emph{lac} operon being either OFF or ON, and each bit of the vector matches exactly what one should expect biologically. The sixth initial condition, when $(L_e,L_{em},G_e)=(0,1,0)$ represents medium levels of extracellular lactose and no glucose. In this case, there are two basins of attractions and two fixed points, one OFF and one ON. This represents the important biological phenomenon of bistability. We will describe this in more details later, because it also arises in our model of the \emph{ara} operon.

\subsection{Proposed Boolean model}

Our model of the \emph{ara} operon consists of nine Boolean variables and four parameters (or constants), which differ from the variables primarily in the time-scales over which they change. For example, the presence of extracellular glucose or arabinose can certainly change, but it might do so on the scale of hours. In contrast, the concentrations of enzymes, proteins, or mRNA may change on a scale of seconds or minutes. Thus, we consider extracellular glucose and arabinose to be parameters of the model, and the concentration of the gene products as variables. We also define a parameter for the unbound AraC protein, since the gene that codes for it is not part of the operon. Our model uses the following Boolean parameters:
\begin{itemize}
  \item $A_e=$ extracellular arabinose (high levels)
  \item $A_{em}=$ extracellular arabinose (at least medium levels)
  \item $Ara_{-}=$ AraC protein (unbound to arabinose)
  \item $G_e=$ extracellular glucose
\end{itemize}
The subscript $m$ denotes medium concentration, and allows us to distinguish between three levels of arabinose concentration. Low or ``basal'' levels occur when $(A_e,A_{em})=(0,0)$, medium levels when $(A_e,A_{em})=(0,1)$, and high levels when $(A_e,A_{em})=(1,1)$. The fourth case, when $(A_e,A_{em})=(1,0)$ can be ignored, as it is meaningless. Next, we define the following Boolean variables:
\begin{itemize}
\item $M_S=$ mRNA of the structural genes ($ara_{BAD}$)
\item $M_T=$ mRNA of the transport genes ($ara_{EFGH}$)
\item $E=$ enzymes AraA, AraB, and AraD, coded for by the structural genes
\item $T=$ transport proteins, coded for by the transport genes
\item $A=$ intracellular arabinose (high levels)
\item $A_m=$ intracellular arabinose (medium levels)
\item $C=$ cAMP-CAP protein complex
\item $L=$ DNA loop
\item $Ara_+=$ arabinose-bound AraC protein
\end{itemize}
The variable $L$ is $1$ if the DNA is looped, and $0$ if it is not looped. All of the other variables represent concentration levels of the corresponding gene product: $1$ for ``high'', and $0$ for ``low''. Note that we distinguish between the AraC protein that is unbound to arabinose (which acts as a repressor) and the arabinose-bound protein (which acts as a promoter), with the parameter $Ara_{-}$ and the variable $Ara_+$, respectively. 

For each of these variables, we propose a Boolean function that represents how its state is influenced by the other variables and parameters. We justify these functions below, using the standard Boolean logic symbols $\And$ for ``AND'', $\Or$ for ``OR'', and $\overline{X}$ to denote ``NOT $X$''. 

\begin{itemize}

\item For the structural genes' mRNA to be transcribed, we need the
  presence of the cAMP-CAP protein complex and the arabinose-bound
  AraC protein, as well as the DNA to be unlooped. The Boolean function is $f_{M_S}=C\And Ara_+\And\overline{L}$.

\item For the transport genes' mRNA to be transcribed, the cAMP-CAP protein complex and the arabinose-bound AraC protein are needed. The Boolean function is $f_{M_T}=C\And Ara_+$.

\item For the enzymes involved in the metabolism of arabinose (AraA,
  AraB, and AraD) to be present, the structural genes' mRNA is
  needed. The Boolean function is $f_E=M_S$.

\item For the transport proteins to be translated, the transport
  genes' mRNA is needed. The Boolean function is $f_T=M_T$.

\item For intracellular arabinose to be at a high concentration,
  high levels of extracellular arabinose and the transport protein must be
  present. The Boolean function is $f_A=A_e\And T$.

\item For intracellular arabinose to be at medium concentration (or
  higher), we require the presence of either extracellular arabinose
  at a high concentration, or a medium concentration of extracellular
  arabinose and the transport protein. The Boolean function is
  $f_{A_m}=(A_{em}\And T) \Or A_e$.

\item For the cAMP-CAP protein complex to be present we require the
  absence of external glucose. The Boolean function is $f_C
  =\overline{G_e}$.

\item For the DNA loop to be formed, the AraC protein must be present, but it cannot be bound to arabinose. The Boolean function is $f_L=Ara_-\And\overline{Ara_+}$.

\item For the arabinose-bound form of the AraC protein to be formed,
  AraC protein must be present, as well as either low or high
  concentrations of intracellular arabinose. The Boolean function
  is $f_{Ara_+}=Ara_-\And(A_m\Or A)$.
\end{itemize}

\begin{figure}
  \tikzstyle{v} = [circle, fill=white,draw,inner sep=0pt, minimum size=6mm] 
  \tikzstyle{w} = [rectangle, fill=white,draw,inner sep=0pt, minimum size=6mm] 
  \tikzstyle{act} = [draw, -stealth]
  \begin{tikzpicture}[scale=1.6]
    \draw[fill=black!20, rounded corners] 
    (-.7,-1.4) rectangle (2.5,2);
    \node[v] (Ms) at (0.8,1.6) {\small $M_S$};
    \node[v] (C) at (-.3,1.6) {\small $C$};
    \node[w] (Ara-) at (0.8,-1) {\small $Ara_-$};
    \node[v] (T) at (2,0.8) {\small $T$};
    \node[v] (E) at (-.3,0.9) {\small $E$};
    \node[v] (L) at (-.3,-.1) {\small $L$};
    \node[v] (Mt) at (2,1.6) {\small $M_T$};
    \node[v] (A) at (2,-.1) {\small $A$};
    \node[v] (Ara+) at (0.8,-.1) {\small $Ara_+$};
    \node[w] (Ge) at (-1.2,1.6) {\small $G_e$};
    \node[w] (Ae) at (3,-.1) {\small $A_e$};
    \draw[inhib] (Ge) to (C);
    \draw[activ] (Ara-) to[bend left=24] (L);
    \draw[activ] (Ara-) to (Ara+);
    \draw[activ] (T) to (A);
    \draw[inhib] (L) to (Ms);
    \draw[activ] (A) to (Ara+);   
    \draw[activ] (Ae) to (A);
    \draw[activ] (C) to[bend left=22] (Mt);
    \draw[activ] (C) to (Ms);
    \draw[activ] (Mt) to (T);
    \draw[activ] (Ms) to (E);
    \draw[inhib] (Ara+) to (L);
    \draw[activ] (Ara+) to (Ms);   
    \draw[activ] (Ara+) to (Mt);
    \begin{scope}[shift={(4.5,.3)}]
      \node[anchor=west] at (0,1.6) {$f_A=A_e \And T$ };
      \node[anchor=west] at (0,1.2) {$f_{A_m}=(A_{em} \And T) \Or A_e$};
      \node[anchor=west] at (0,.8) {$f_{Ara_+}=(A_m \Or A)\And Ara_-$};
      \node[anchor=west] at (0,.4) {$f_C=\overline{G_e}$ };
      \node[anchor=west] at (0,0) {$f_E=M_S$};
      \node[anchor=west] at (0,-.4) {$f_L=\overline{Ara_+}\And Ara_-$};
      \node[anchor=west] at (0,-.8) {$f_{M_S}=Ara_+\And C\And \overline{L}$};
      \node[anchor=west] at (0,-1.2) {$f_{M_T}=Ara_+ \And C$ };
      \node[anchor=west] at (0,-1.6) {$f_T=M_T$};
      \draw[decorate,decoration={brace,amplitude=5pt}] 
      (0,-1.75) --  (0,1.75); 
    \end{scope}
  \end{tikzpicture}
  \caption{The wiring diagram (left) and logical functions (right) of
    our proposed Boolean model of the \emph{ara} operon. Circles
    represent variables and rectangles represent parameters. The two
    nodes for intracellular arabinose ($A$ and $A_m$) are collapsed
    into one, as are the two nodes for extracellular arabinose.}
\end{figure}
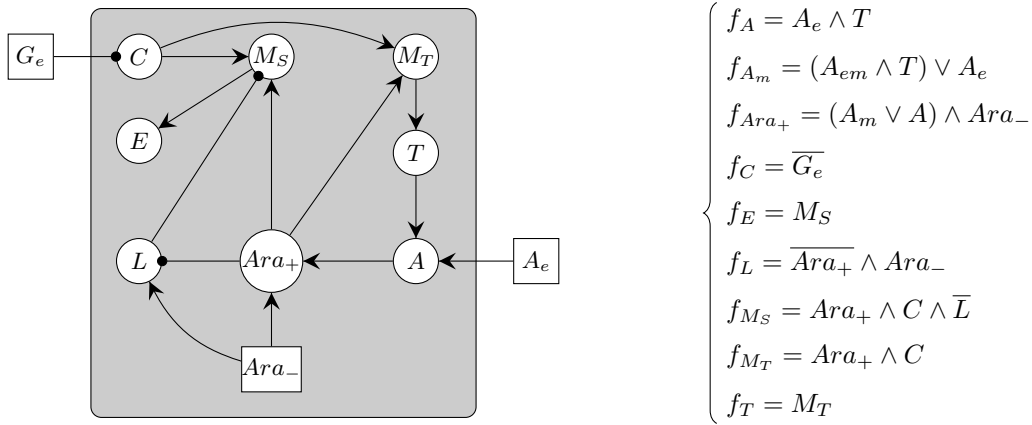

Those readers familiar with the Boolean model of the \emph{lac} operon in \cite{veliz-cuba2011boolean} may notice that we do not include a $\And\overline{G_e}$ clause in $f_A$ or $f_{A_m}$ like Veliz-Cuba and Stigler do in the functions $f_L$ and $f_{L_m}$ for intracellular lactose. The lactose and arabinose transporter proteins are different, and glucose does not repress arabinose via inducer exclusion the way it does lactose.

\section{Analysis and model validation}\label{sec:analysis}

There are $2^4=16$ possibilities for the parameter vector $x=(A_e,A_{em},G_e,Ara_{-})\in\F_2^4$, but we disregard the four where $A_e=1$ and $A_{em}=0$ because they are meaningless. This leaves $12$ possible initial conditions of our model, and we need to consider all of these in our analysis of the system dynamics. For each one, we look at the periodic orbit structure of the phase space. Naturally, this depends on the update scheme. However, in any Boolean network, the fixed points are the same under synchronous and asynchronous update. In this section, we use computational algebra to compute the fixed points for all $12$ initial conditions. For all $11$ cases of a single fixed point, we argue why that steady state makes sense biologically. After that, we examine the case with two fixed points that predicts bistability. Finally, we need to verify that there are no larger periodic orbit structures, and for this we turn to the GINsim (Gene Interaction Network simulation) software. It turns out that longer limit cycles appear under a synchronous update order, but these go away for general asynchronous update.

\subsection{Fixed points and Gr\"obner bases}

The fixed points of our Boolean network model can be found by solving the system
$\{f_{x_i} = x_i\mid i = 1,\dots,9\}$ where we rename our Boolean
variables as follows:
\[
(A, A_m, Ara_{+}, C, E, L, M_S, M_T, T) = (x_1, x_2, x_3, x_4, x_5, x_6, x_7, x_8, x_9).
\]
The resulting system is shown below, in Boolean logical form on the left, and in polynomial form on the right. The conversion from logical to polynomial form is done by replacing $x\And y$ with $xy$, $x\Or y$ with $x+y+xy$, and $\overline{x}$ with $1+x$. 
\[
  \tikzstyle{v} = [circle, fill=white,draw,inner sep=0pt, minimum size=6mm] 
  \tikzstyle{w} = [rectangle, fill=white,draw,inner sep=0pt, minimum size=6mm] 
  \begin{tikzpicture}[scale=1.5]
    \begin{scope}[shift={(0,0)}]
      \node[anchor=west] at (0,1.6) {$f_A=A_e \And T=A$ };
      \node[anchor=west] at (0,1.2) {$f_{A_m}=(A_{em} \And T) \Or A_e =A_m$};
      \node[anchor=west] at (0,.8) {$f_{Ara_+}=(A_m \Or A)\And Ara_{-}=Ara_+$ };
      \node[anchor=west] at (0,.4) {$f_C=\overline{G_e}=C$ };
      \node[anchor=west] at (0,0) {$f_E=M_S=E$};
      \node[anchor=west] at (0,-.4) {$f_L=\overline{Ara_+}\And Ara_-=L$ };
      \node[anchor=west] at (0,-.8) {$f_{M_S}=Ara_+\And C\And \overline{L}=M_S$ };
      \node[anchor=west] at (0,-1.2) {$f_{M_T}=Ara_+ \And C=M_T$ };
      \node[anchor=west] at (0,-1.6) {$f_T=M_T=T$};
      \draw[decorate,decoration={brace,amplitude=5pt}] 
      (0,-1.8) --  (0,1.8); 
      \node at (4.15,0) {$\Longleftrightarrow$};
    \end{scope}
    \begin{scope}[shift={(5,0)}]
      \node[anchor=west] at (0,1.6) {$x_1+x_9A_e=0$ };
      \node[anchor=west] at (0,1.2) {$x_2+A_{em}x_9+A_e+A_{em}A_ex_9=0$ };
      \node[anchor=west] at (0,.8) {$x_3+(x_1+x_2+x_1x_2)Ara_{-}=0$ };
      \node[anchor=west] at (0,.4) {$x_4+G_e+1=0$ };
      \node[anchor=west] at (0,0) {$x_5+x_7=0$};
      \node[anchor=west] at (0,-.4) {$x_6+(x_3+1)Ara_{-}=0$ };
      \node[anchor=west] at (0,-.8) {$x_7+x_3x_4(x_6+1)=0$ };
      \node[anchor=west] at (0,-1.2) {$x_8+x_3x_4=0$ };
      \node[anchor=west] at (0,-1.6) {$x_9+x_8=0$};
      \draw[decorate,decoration={brace,amplitude=5pt}] 
      (0,-1.8) --  (0,1.8); 
    \end{scope}
  \end{tikzpicture}
\]
Systems such as these can be solved rather easily by computing a Gr\"obner basis with computational algebra software such as Macaulay2 or Sage. As an example, when the parameter vector is $(A_e,A_{em},Ara_{-},G_e)=(0,1,1,0)$, we type the following commands into Sage:
\begin{verbatim}
  P.<x1,x2,x3,x4,x5,x6,x7,x8,x9> = PolynomialRing(GF(2),9,order='lex');
  Ae = 0; Aem = 1; ara=1; Ge = 0;
  I = ideal(x1+x9*Ae, x2+(Aem*x9+Ae+Aem*Ae*x9), x3+(x1+x2+x1*x2)*ara, 
      x4+Ge+1, x5+x7, x6+(x3+1)*ara, x7+x3*x4*(x6+1), x8+x3*x4, x9+x8);
  B = I.groebner.basis();
\end{verbatim}
The output is the following:

{\small
\begin{verbatim}
  [x1, x2 + x9, x3 + x9, x4 + 1, x5 + x9^2, x6 + x9 + 1, x7 + x9^2, x8 + x9].
\end{verbatim} 
} This says that the original nonlinear system of equations has the same set of solutions over $\F_2$ as the much simpler system 
\[
\{x_1=0,x_2+x_9=0,x_3+x_9=0,x_4+1=0,x_5+x_9^2=0,x_6+x_9+1=0,x_7+x_9^2=0,x_8+x_9=0\}. 
\]
It is easy to verify by hand (recall that $x_i^2=x_i$) that there are two solutions:
\[
x_{OFF}=(0,0,0,1,0,1,0,0,0),\qquad\text{and}\qquad
x_{ON}=(0,1,1,1,1,0,1,1,1).
\]
We will now analyze and interpret these fixed points. The parameter vector $(A_e,A_{em},Ara_{-},G_e)=(0,1,1,0)$ means that there are medium levels of extracellular arabinose but no glucose, and that the unbound AraC protein is present. The solution $x_{OFF}$ corresponds to the steady state when the cAMP-CAP complex is present $(C=1)$, the DNA is looped $(L=1)$, and all other variables are $0$. This makes sense biologically, since the DNA loop blocks transcription, thereby preventing the other gene products from being present: arabinose, the arabinose-bound AraC protein, mRNA, and the enzymes and transporter proteins. In this case, the operon is OFF. The other steady state $x_{ON}$ corresponds to the case where the arabinose-bound AraC protein and all of the gene products are present, but the DNA loop is not ($L=0$), and internal arabinose levels are medium by not high ($A_m=1$, $A=0$). This describes the \emph{ara} operon being ON. 

The fact that there are two fixed points for the initial condition described above shows how the operon predicts bistability, and this will be discussed in more detail in Section~\ref{subsec:bistability}. Before that, there are $11$ other initial conditions that need to be analyzed in a similar manner. In all of these cases, there is only one fixed point. These are summarized in Table~\ref{tbl:fixed-points}, grouped by the fixed point. We will justify the biological interpretation of these fixed points now, going down the table from top-to-bottom.

\begin{table}
\begin{tabular}{|c|c|c|}
\hline  Initial condition(s) & Fixed point(s) & Operon \\ 
 $x=(A_e,A_{em},Ara_{-},G_e)$ & 
$(A,A_m,Ara_{+},C,E,L,M_S,M_T,T)$ & state \\
\hline 

$(0,0,0,0)$ & $(0,0,0,1,0,0,0,0,0)$ & OFF \\
$(0,1,0,0)$ & & \\ \hline 

$(1,1,0,0)$ & $(0,1,0,1,0,0,0,0,0)$ & OFF \\ \hline 

$(0,0,0,1)$ & $(0,0,0,0,0,0,0,0,0)$ & OFF \\
$(0,1,0,1)$ & & \\ \hline

$(1,1,0,1)$ & $(0,1,0,0,0,0,0,0,0)$ & OFF \\ \hline

$(0,0,1,0)$ & $(0,0,0,1,0,1,0,0,0)$ & OFF \\ \hline 

$(0,0,1,1)$ & $(0,0,0,0,0,1,0,0,0)$ & OFF \\
$(0,1,1,1)$ & & \\ \hline
$(1,1,1,1)$ & $(0,1,1,0,0,0,0,0,0)$ & OFF \\ \hline 

$(1,1,1,0)$ & $(1,1,1,1,1,0,1,1,1)$ & ON \\ \hline 

$(0,1,1,0)$ & $(0,0,0,1,0,1,0,0,0)$ & OFF \\
 & $(0,1,1,1,1,0,1,1,1)$ & ON \\ \hline
\end{tabular}
\caption{The fixed points of our Boolean network model for each choice of parameters, excluding the nonsensical ones where $A_e=1$ and $A_{em}=0$.} \label{tbl:fixed-points}
\end{table}

We start with the two initial conditions for which $A_e=Ara_{-}=G_e=0$, which represent the cases of at most medium concentration levels of extracellular arabinose, no glucose, and no AraC protein (e.g., due to a mutation in the \emph{araC} gene). As shown in the first row of Table~\ref{tbl:fixed-points}, the fixed point of this system has only $C=1$. This makes biological sense due to the following arguments: arabinose does not enter the cell (so $A=A_m=0$). Thus, there is no AraC protein bound to arabinose (so $Ara_+=0$), which means that the DNA is not looped $(L=0)$. Despite the absence of the DNA loop which blocks transcription, the absence of the AraC protein prohibits the initiation of transcription, so all of the gene products will be absent ($E=M_S=M_T=T=0$). Finally, the cAMP-CAP protein complex will be present ($C=1$) because glucose is needed to reduce the concentration of CAP.

A similar case holds for the initial condition $(A_e,A_{em},Ara_{-},G_e)=(1,1,0,0)$. Here, the only difference is that the high level of extracellular arabinose results in a few arabinose molecules entering the cell via passive transport, so $A_m=1$. The rest of the argument is the same.

Another case similar to the first two occurs for the three initial conditions for which $Ara_-=0$ and $G_e=1$. Once again, the first argument above carries through, except that the presence of glucose reduces CAP concentration levels. This results in $C=0$ and the zero vector as the steady state, unless there are high levels of extracellular arabinose, in which case $A_m=1$ due to passive transport.

The remaining cases all describe initial conditions where the AraC protein is present, i.e., $Ara_-=1$. For the first, there is no arabinose or glucose, and so the DNA will be looped and cAMP-CAP will be high. The remaining variables should be zero because transcription cannot initiate with $Ara_+=0$. In the next two cases, glucose is present, so the cAMP-CAP complex will not be, and transcription cannot begin. This means that there will be no transporter proteins, and since the extracellular arabinose levels are not high, the intracellular arabinose levels will remain low. All other variables will be $0$. 

Finally, consider the two initial conditions with high levels of extracellular arabinose and $Ara_-=1$. Some arabinose will enter the cell via passive transport and bind to the AraC protein, so $Ara_+=1$ and the DNA will be unlooped. If glucose is present, then CAP levels are low and so transcription cannot begin, and the fixed point that is reached is OFF, with only $A_m=Ara_+=1$. If glucose is absent, then transcription initiates, and the ON fixed point is reached, with all states $1$ except $L=0$.

\subsection{Bistability}\label{subsec:bistability}

\begin{figure}
\includegraphics[width=.8\textwidth]{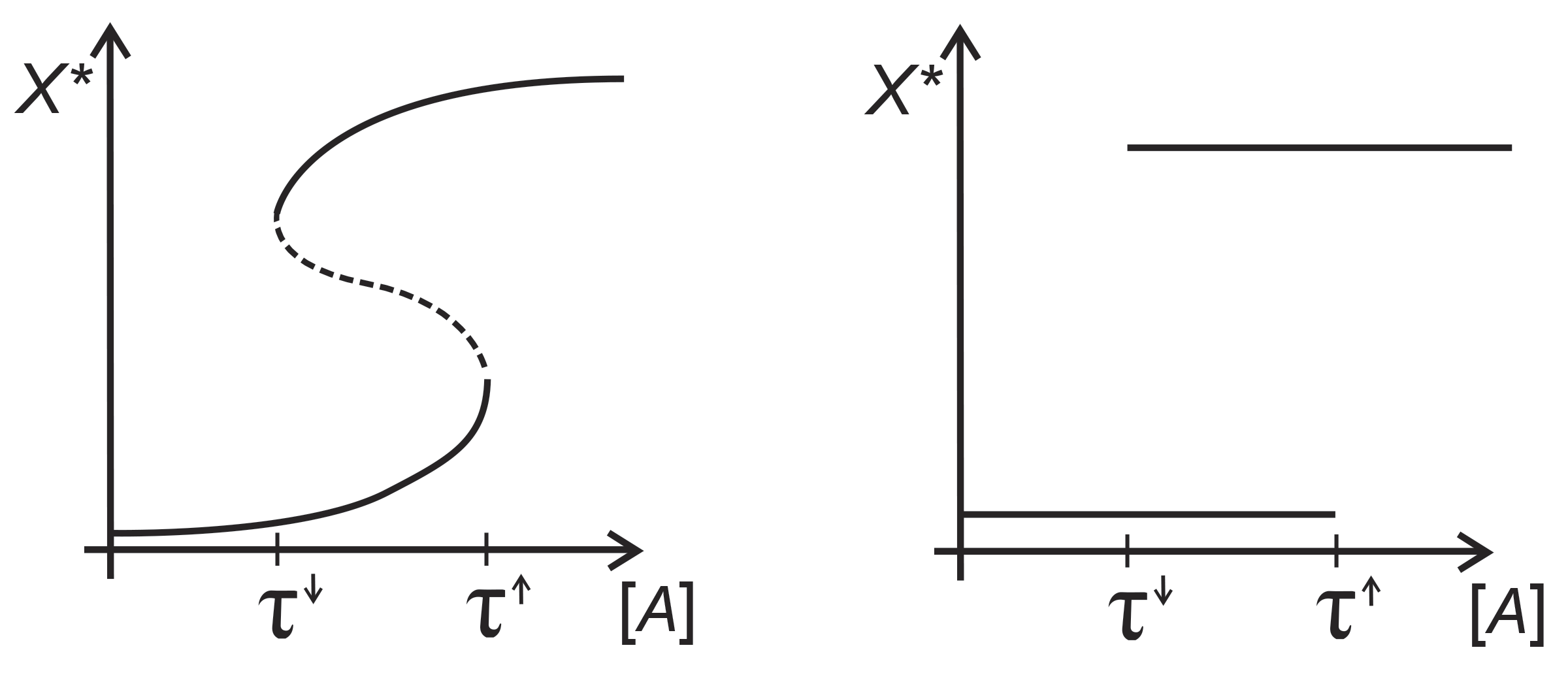}
\caption{On the left is a curve the describes the steady states of a gene product $X$ in an ODE model, as a function of concentration of extracellular arabinose. On the right is a Boolean analogue of this. The dashed curve represents unstable steady states. Bistability occurs when $[A]\in(\tau^\downarrow,\tau^\uparrow)$ because there are two stable steady states.}\label{fig:bistability}
\end{figure}

Let us return to the initial condition describing medium levels of extracellular arabinose and no glucose. Biologically, this captures the phenomenon of \emph{bistability}, which says that the operon can reside in either the OFF or ON steady state. In classical differential equation models, bistability occurs when there are two steady states which are separated by an unstable steady state. To understand this in the \emph{ara} operon, let $[A]$ denote concentration of intracellular arabinose, and fix a gene product $X$ such as mRNA, an enzyme, or protein, with concentration $[X]$. If $[A]\approx 0$, then there is one steady state solution $[X]=X^*\approx 0$. One wishes to understand how the steady state $X^*$ depends on $[A]$, and this can be plotted in the $([A],X^*)$-plane. An example of this that could arise from a differential equation model is shown on the left in Figure~\ref{fig:bistability}. If $[A]\approx 0$, then $X^*\approx 0$ as well. As $[A]$ increases continuously, then so does $X^*$, until $[A]$ exceeds some threshold $\tau^\uparrow$, at which point $X^*$ undergoes a discontinuous ``jump'' from the operon being OFF to ON. However, once the operon is in the ON state, $[A]$ must be reduced to some lower threshold $\tau^\downarrow<\tau^\uparrow$ before $X^*$ can ``jump'' down to the OFF state. This means that if the arabinose concentration $[A]$ is in the range $(\tau^\downarrow,\tau^\uparrow)$, then the operon can be observed as ON in some cells and OFF in others. Transcription of the \emph{araBAD} genes will likely not be initiated in cells that were raised in an arabinose-starved environment when $[A]$ gets increased to this middle range. In contrast, transcription will likely be observed in cells raised with arabinose, even as $[A]$ is decreased into this range. The S-shape of the curve in Figure~\ref{fig:bistability} is due to a third unstable steady state in the bistable region, which is denoted by the dashed line. The Boolean analogue of this bistability is shown on the right in Figure~\ref{fig:bistability}. 

To see how our Boolean network captures bistability, consider the initial condition whose parameter vector is $x=(A_e,A_{em},Ara_-,G_e)=(0,1,1,0)$ and look at Table~\ref{tbl:fixed-points}. Some cells, such as those raised in an arabinose-starved environment, end up in the OFF fixed point $(0,0,0,1,0,1,0,0,0)$. Now, consider increasing extracellular arabinose levels. When the concentration exceeds the ``up-threshold'' $\tau^\uparrow$, then $A_e$ flips to $1$ in the parameter vector. This results in the system settling in a new ON fixed point $(1,1,1,1,1,0,1,1,1)$. On the other hand, cells raised in an arabinose-rich environment will be initially observed in the ON fixed point $(0,1,1,1,1,0,1,1,1)$. But if arabinose levels get too low, then $A_{em}=0$ in the parameter vector, and the system settles in a new OFF fixed point, $(0,0,0,1,0,1,0,0,0)$. 

\subsection{Longer limit cycles and general asynchronous update}

At this point, we have verified that the fixed points of our Boolean network model make sense biologically for all 12 choices of initial conditions. It remains to check that there are no other longer periodic cycles in the phase space. It would be nice to have algebraic methods to do this or a theorem with sufficient conditions for having only one basin of attraction. As far as we know, neither of these exist, and so we need to resort to computer simulations. Using the freely available online software \emph{Analysis of Dynamic Algebraic Models} (ADAM)\footnote{The ADAM software is being overhauled and replaced with a crowd-sourced version called TURING, currently in Beta \cite{turing}.} \cite{hinkelmann2011adam}, we verified that indeed, for the $11$ initial conditions with only fixed points, there is only one basin of attraction consisting of all $512$ nodes. However, for the case where bistability is observed is more complicated. The phase space actually consists of six attractor basins: two with $24$ nodes each that lead into the fixed points, one with $80$ nodes that leads into a $2$-cycle, $(0,0,1,1,0,1,0,0,1)\longto(0,1,0,1,0,0,0,1,0)$, and three with $128$ nodes each, that lead into the following $4$-cycles:
\[
(0,1,0,1,0,1,0,0,0)\longto (0,0,1,1,0,1,0,0,0)\longto (0,0,0,1,0,0,0,1,0)\longto(0,0,0,1,0,1,0,0,1),
\]
\[
(0,1,0,1,1,1,0,0,1)\longto (0,1,1,1,0,1,0,0,0)\longto (0,0,1,1,0,0,0,1,0)\longto(0,0,0,1,0,0,1,1,1),
\]
\[
(0,0,1,1,0,0,1,1,1)\longto (0,1,0,1,1,0,1,1,1)\longto (0,1,1,1,1,1,0,0,1)\longto(0,1,1,1,0,0,0,1,0),
\]
This presents a problem because these longer periodic cycles do not make sense biologically. One explanation could be that that the model is faulty, in that either it does not represent the biology well, or perhaps the \emph{ara} operon is simply ill-suited for a Boolean network framework. Another possibility could be that these periodic cycles are an artifact of using a synchronous update order. To test this, we used the \emph{Gene Interaction Network simulation} (GINsim) software \cite{chaouiya2012logical} to analyze the phase space under a general asynchronous update, which is more representative of an actual molecular network. Under a general asynchronous update, the periodic cycles merge and the only attractors in the phase space are the original two fixed points. This provides evidence that the longer periodic cycles are indeed artifacts of a synchronous update order and not a flaw in the model design.

\section{Discussion and future work}\label{sec:conclusions}

In this paper, we proposed a Boolean network model of the \emph{ara} operon in E.~coli on $9$ variables and $4$ parameters. Unlike the \emph{lac} operon which was modeled with a Boolean network in \cite{veliz-cuba2011boolean}, the \emph{ara} operon uses DNA looping to block transcription, it has a protein that acts as both a repressor and an activator, and glucose does not repress the transport of arabinose via inducer exclusion. We showed that under 11 of the 12 possible initial conditions, the system reaches a fixed point that agrees with the biology. In the last case, there were two fixed points -- one for the operon being ON and the other OFF. This captures the notion of bistability under medium levels of arabinose and no glucose. Though there were other periodic cycles under an ``artificial'' synchronous update, these went away under a general asynchronous update order. The chance that this would happen in all of these cases by pure luck is astronomically small, which provides strong evidence for the strength and robustness of our model and the Boolean network framework.

Our analysis used both computational algebra and software tools. We found the fixed points by writing the Boolean functions as polynomials and then computing a Gr\"obner basis using Sage. Of course, this was not necessary because this network is small enough that we could have used Boolean network software such as ADAM and/or GINsim to complete it. However, these methods are necessary for larger networks where the phase space is too big for Boolean network software to handle. That said, one of the shortcomings is that not all of the model analysis can be done using computational algebra, and it would be desirable to further develop this theory. For example, is there a way to deduce that that a model has only one attractor basin, or only fixed points, just from the Boolean functions? There are some known results and theorems along these lines, but we were not able to find any the directly applied to our model. Examples of such results include the following, mostly for general asynchronous update. In \cite{robert1980iterations}, Robert showed that if the wiring diagram has no directed cycle, then $f$ has a unique fixed point. However, this does not eliminate the possibility of other periodic cycles or complex attractors. Robert's result was generalized by Shih and Dong by restricting the directed cycles to certain subgraphs of the wiring diagram \cite{shih2005combinatorial}. In \cite{richard2015fixed}, Richard gave a sufficient condition by proving a ``forbidden subnetworks theorem.'' By a theorem of Richard \cite{richard2010negative}, negative cycles are necessary for the existence of a cyclic attractor. Since the \emph{ara} operon has no negative cycles, the phase space cannot contain periodic cycles. However, that does not necessarily forbid the existence of long complex attractors. Another body of work involves the analysis and control of Boolean networks using semi-tensor products \cite{cheng2011analysis}, but it is not clear whether this work is directly applicable. Future work on mathematical side includes either finding or developing theorems that would apply for this type of system. Current and future work on the biological side consists of modeling not only other operons in the Boolean network framework, but more complicated systems such as regulons and modulons. In some cases, if the biology is not completely understood, then algebraic reverse engineering techniques (e.g., \cite{laubenbacher2004computational} or \cite{dimitrova2011parameter}) can predict interactions are dependencies that have not been observed biologically.


\bibliographystyle{alpha}
\newcommand{\etalchar}[1]{$^{#1}$}


\end{document}